\documentclass[12pt]{article}
\usepackage[dvips]{graphicx}
\usepackage{epsfig}
\usepackage{amsmath,amssymb,amsbsy}
\usepackage{amsfonts}
\begin{document}
\thispagestyle{empty}
\begin{center}

{\Large\bf{$W^{\pm}$ bosons production
in the quantum statistical parton distributions approach}}

\vskip1.4cm
{\bf Claude Bourrely}
\vskip 0.3cm
Aix-Marseille Universit\'e, D\'epartement de Physique, \\
Facult\'e des Sciences de Luminy, 13288 Marseille, Cedex 09, France\\
\vskip 0.5cm
{\bf Franco Buccella}
\vskip 0.3cm
INFN, Sezione di Napoli, Via Cintia, I-80126 Napoli, Italy
\vskip 0.5cm
{\bf Jacques Soffer}
\vskip 0.3cm
Physics Department, Temple University\\
Barton Hall, 1900 N, 13th Street\\
Philadelphia, PA 19122-6082, USA
\vskip 0.5cm
{\bf Abstract}\end{center}

We consider $W^{\pm}$ gauge bosons production in connection with recent results from BNL-RHIC and FNAL-Tevatron and interesting predictions from the statistical parton distributions. They concern relevant aspects of the structure of the nucleon sea and the high-$x$ region of the valence quark distributions.
We also give predictions in view of future proton-neutron collisions experiments at BNL-RHIC.

\vskip 0.5cm


\vskip 0.5cm

\newpage
\section{Introduction}
 Several years ago a new set of parton distribution functions (PDF) was constructed in the
framework of a statistical approach of the nucleon \cite{bbs1}, and we
first recall very briefly, its main characteristic features. For quarks
(antiquarks), the building blocks are the helicity dependent distributions
$q^{\pm}(x)$ ($\bar q^{\pm}(x)$). This allows to describe simultaneously the
unpolarized distributions $q(x)= q^{+}(x)+q^{-}(x)$ and the helicity
distributions $\Delta q(x) = q^{+}(x)-q^{-}(x)$ (similarly for antiquarks). At
the initial energy scale taken at $Q^2_0= 4 \mbox{GeV}^2$, these distributions
are given by the sum of two terms, a quasi Fermi-Dirac function and a helicity
independent diffractive
contribution, which leads to a universal behavior at very low $x$ for all
flavors. The flavor asymmetry for the light sea, {\it i.e.} $\bar d (x) > \bar
u (x)$, observed in the data is built in. This is clearly understood in terms
of the Pauli exclusion principle, based on the fact that the proton contains
two up-quarks and only one down-quark. The chiral properties of QCD lead to
strong relations between $q(x)$ and $\bar q (x)$.
For example, it is found that the well estalished result $\Delta u (x)>0 $\
implies $\Delta
\bar u (x)>0$ and similarly $\Delta d (x)<0$ leads to $\Delta \bar d (x)<0$. 
This earlier prediction was confirmed by recent data. Concerning
 the gluon, the unpolarized distribution $G(x,Q_0^2)$ is given in
terms of a quasi Bose-Einstein function, with only {\it one free parameter},
and for consistency, one assumes zero gluon polarization, {\it i.e.} $\Delta
G(x,Q_0^2)=0$, at the initial energy scale $Q_0^2$. All unpolarized and
helicity light quark distributions depend upon {\it eight}
free parameters, which were determined in 2002 (see Ref.~\cite{bbs1}), from a
next-to-leading (NLO) fit of a selected set of accurate DIS data. Concerning
the strange quarks and antiquarks distributions, the statistical approach has
been applied to calculate the strange quark asymmetry and the corresponding
helicity distributions, which were found both negative at all $x$ values
\cite{bbs2}. Since the first determination of the free parameters, new tests against experimental (unpolarized and
polarized) data turned out to be very satisfactory, in particular in hadronic
reactions, as reported in Refs.~\cite{bbs3,bbs4}.\\
In this letter, we will consider some physical observables related to $W^{\pm}$ boson production, in connection
with specific features of the statistical distributions in the light quark sector, in particular the 
charge asymmetry  and the parity-violating helicity asymmetries. This is another case where we are able to treat simultaneously
unpolarized and helicity distributions.

\section{The statistical parton distributions}
We now review some of the basic features of the statistical approach, as oppose
to the standard polynomial type
parametrizations of the PDF, based on Regge theory at low $x$ and on counting
rules at large $x$.
The fermion distributions are given by the sum of two terms \cite{bbs1},
a quasi Fermi-Dirac function and a helicity independent diffractive
contribution equal for all light quarks:
\begin{equation}
xq^h(x,Q^2_0)=
\frac{AX^h_{0q}x^b}{\exp [(x-X^h_{0q})/\bar{x}]+1}+
\frac{\tilde{A}x^{\tilde{b}}}{\exp(x/\bar{x})+1}~,
\label{eq1}
\end{equation}
\begin{equation}
x\bar{q}^h(x,Q^2_0)=
\frac{{\bar A}(X^{-h}_{0q})^{-1}x^{2b}}{\exp [(x+X^{-h}_{0q})/\bar{x}]+1}+
\frac{\tilde{A}x^{\tilde{b}}}{\exp(x/\bar{x})+1}~,
\label{eq2}
\end{equation}
at the input energy scale $Q_0^2=4 \mbox{GeV}^2$.\\
The parameter $\bar{x}$ plays the role of a {\it universal temperature}
and $X^{\pm}_{0q}$ are the two {\it thermodynamical potentials} of the quark
$q$, with helicity $h=\pm$. Notice the change of sign of the potentials
and helicity for the antiquarks \footnote{~At variance with statistical mechanics where the distributions are expressed in terms of the energy, here one uses
 $x$ which is clearly the natural variable entering in all the sum rules of the parton model.}.\\
For $q=u,d$, the {\it eight} free parameters\footnote{~We have $A=1.74938$ and $\bar{A}~=1.90801$, which are
fixed by the following normalization conditions $u-\bar{u}=2$, $d-\bar{d}=1$.}
in Eqs.~(\ref{eq1},\ref{eq2}), were
determined at the input scale from the comparison with a selected set of
very precise unpolarized and polarized DIS data \cite{bbs1}. They have the
following values
\begin{equation}
\bar{x}=0.09907,~ b=0.40962,~\tilde{b}=-0.25347,~\tilde{A}=0.08318,
\label{eq3}
\end{equation}
\begin{equation}
X^+_{0u}=0.46128,~X^-_{0u}=0.29766,~X^-_{0d}=0.30174,~X^+_{0d}=0.22775~.
\label{eq4}
\end{equation}
For the gluons we consider the black-body inspired expression
\begin{equation}
xG(x,Q^2_0)=
\frac{A_Gx^{b_G}}{\exp(x/\bar{x})-1}~,
\label{eq5}
\end{equation}
a quasi Bose-Einstein function, with $b_G=0.90$, the only free parameter, since $A_G=20.53$ is determined by the momentum sum
rule.
 For consistency, we also assume that, at the input energy scale, the polarized gluon,
distribution vanishes, so
\begin{equation}
x\Delta G(x,Q^2_0)=0~.
\label{eq6}
\end{equation}
For the strange quark distributions, the simple choice made in Ref. \cite{bbs1}
was greatly improved in Ref. \cite{bbs2}, but they will not be considered in
this paper.\\
In Eqs.~(\ref{eq1},\ref{eq2}) the multiplicative factors $X^{h}_{0q}$ and
$(X^{-h}_{0q})^{-1}$ in
the numerators of the non-diffractive parts of the $q$'s and $\bar{q}$'s
distributions, imply a modification
of the quantum statistical form, we were led to propose in order to agree with
experimental data. The presence of these multiplicative factors was justified
in our earlier attempt to generate the transverse momentum dependence (TMD)
\cite{bbs5}, which was revisited recently \cite{bbs6}.

\section{$W^{\pm}$ production in $\bar{p} p$, $pp$ and $pn$ collisions}
\begin{figure}[htp]
\vspace*{-3.5ex}
\begin{center}
\includegraphics[width=14.0cm]{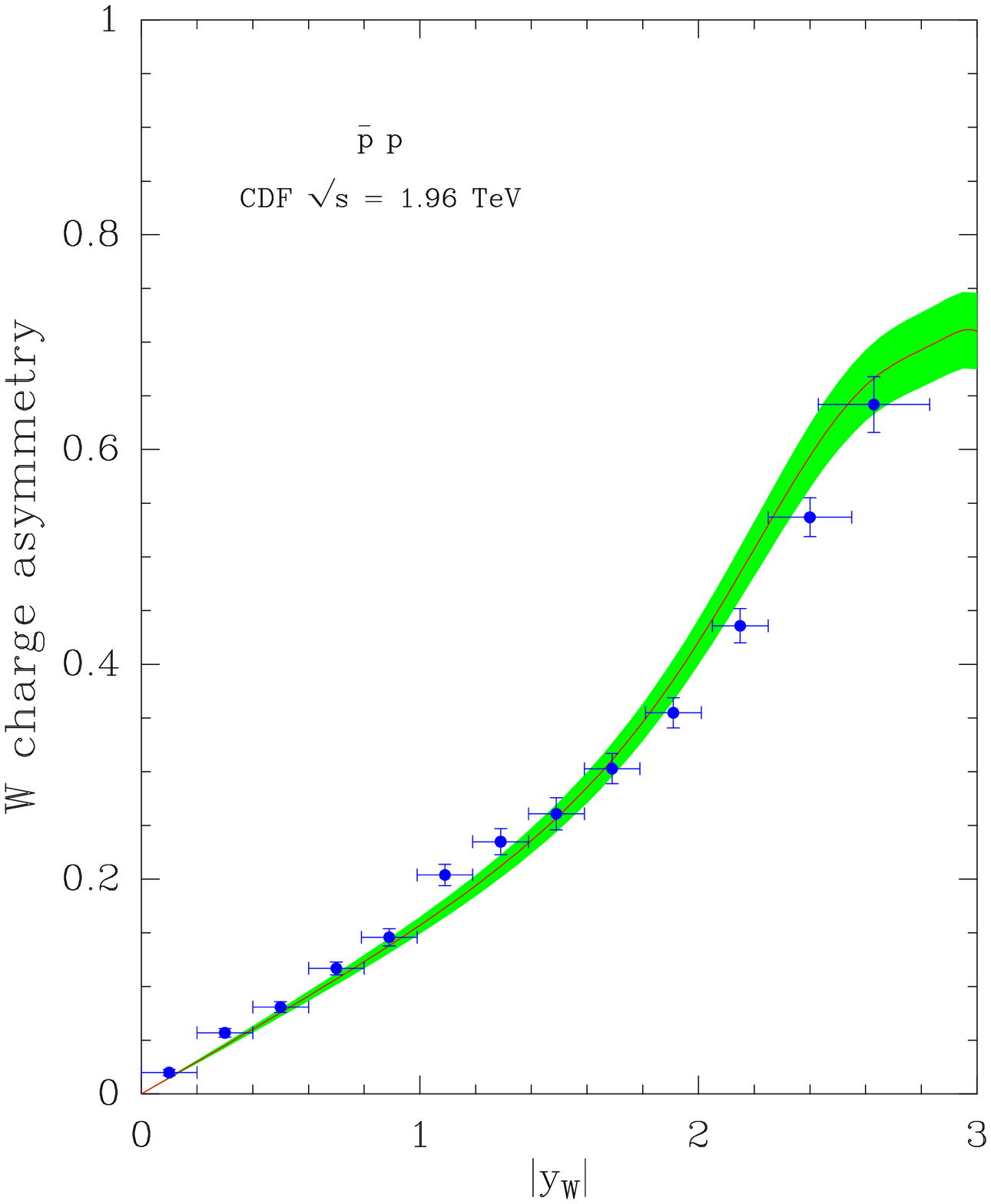}
\caption[*]{\baselineskip 1pt
 The measured W production charge asymmetry from CDF \cite{cdf} versus the $W$ rapidity $y_W$ and the prediction from the statistical approach. The green band represents the uncertainty with a CL of 68{\%}}
\label{asym}
\end{center}
\end{figure}

\begin{figure}[htp]
\vspace*{-3.5ex}
\begin{center}
\includegraphics[width=14.0cm]{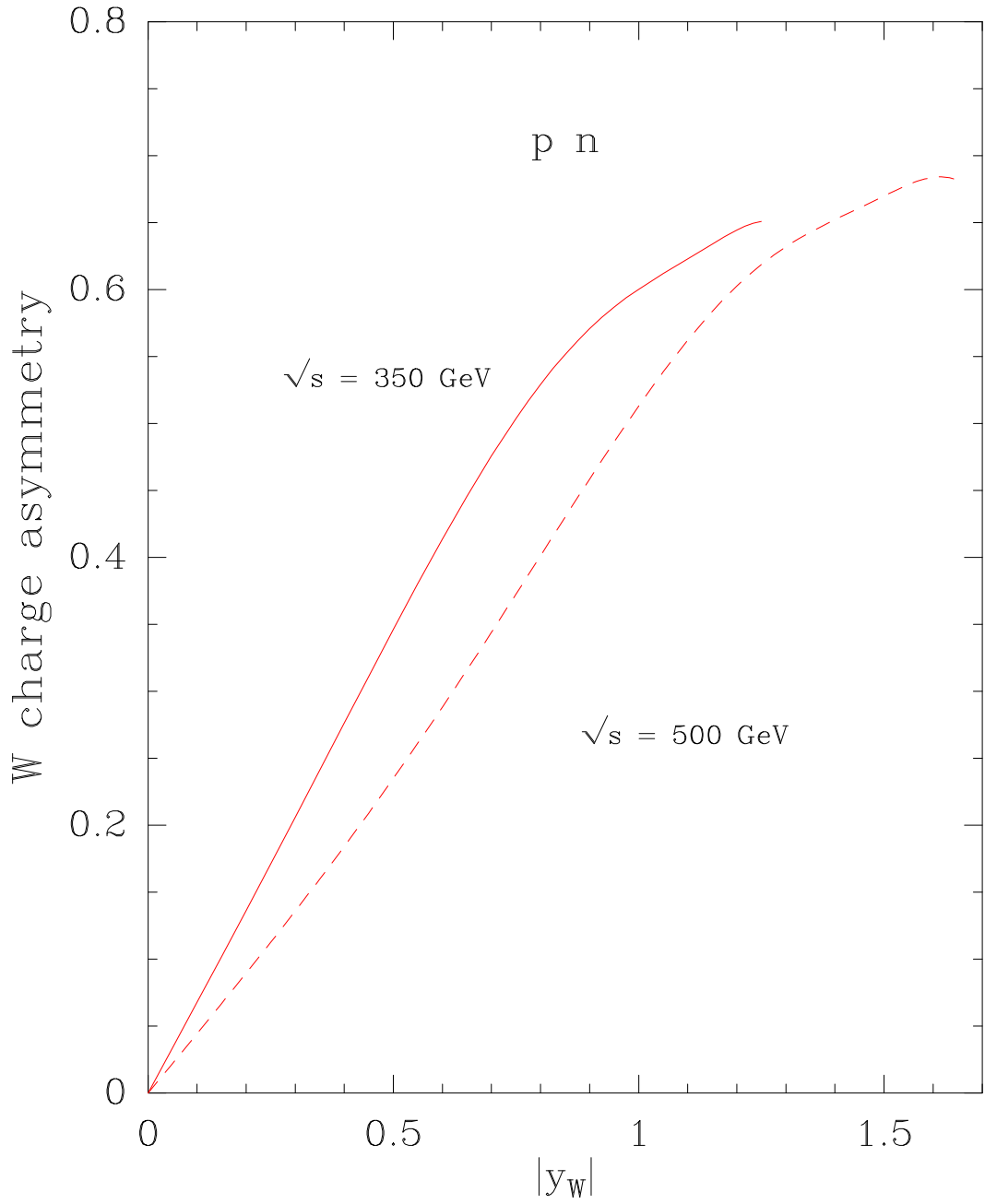}
\caption[*]{\baselineskip 1pt
 Our predicted W production charge asymmetry versus the $W$ rapidity $y_W$, in $pn$ collision calculated at two $c.m.$ energies.}
\label{Rwn}
\end{center}
\end{figure}

\begin{figure}[htp]
\vspace*{-3.5ex}
\begin{center}
\includegraphics[width=14.0cm]{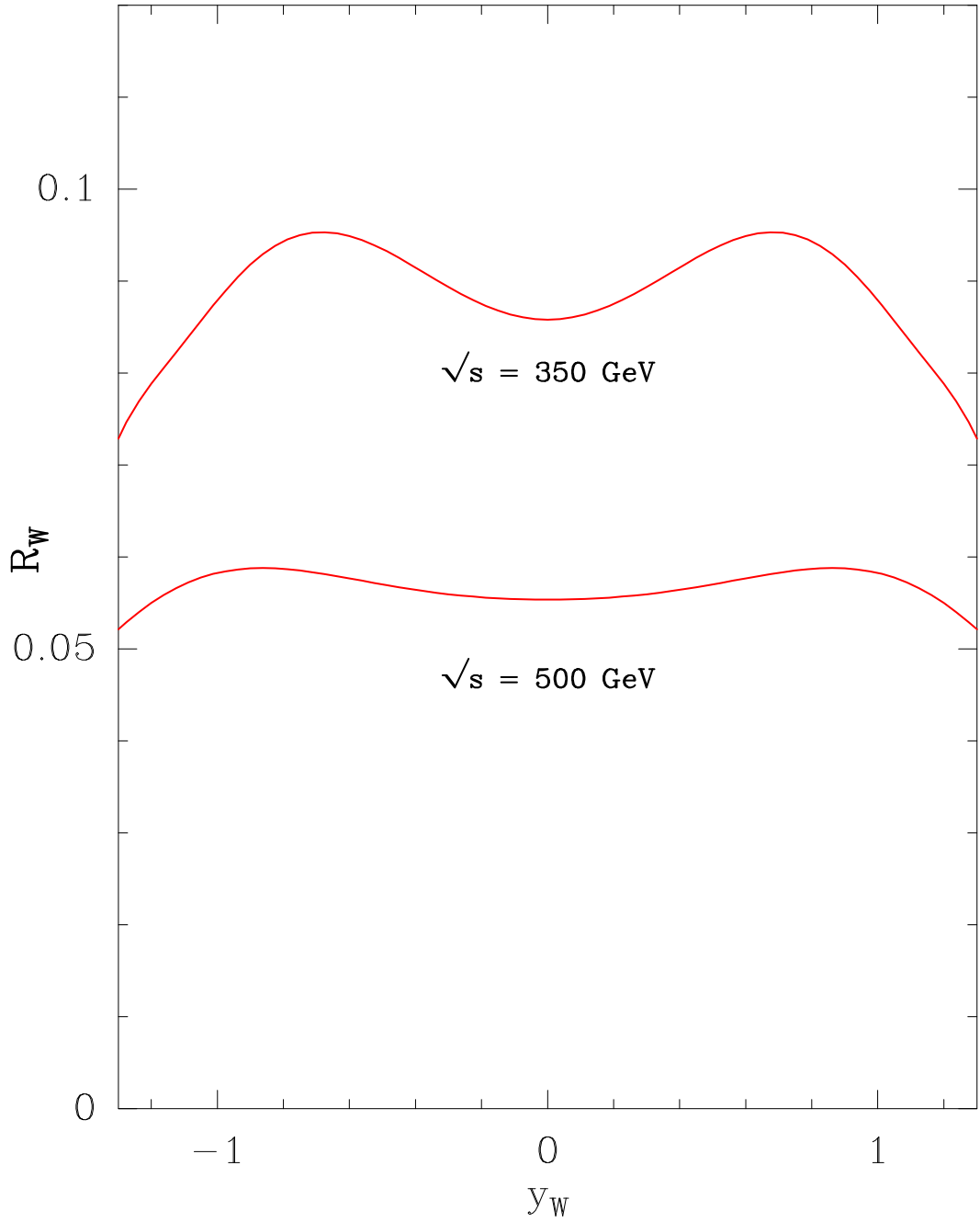}
\caption[*]{\baselineskip 1pt
 Our predicted charge asymmetry $R_W$ involving $pp$ and $pn$ collisions (see Eq. (\ref{Ry1})) calculated at two $c.m.$ energies.}
\label{RW}
\end{center}
\end{figure}

\begin{figure}[htp]
\vspace*{-3.5ex}
\begin{center}
\includegraphics[width=14.0cm]{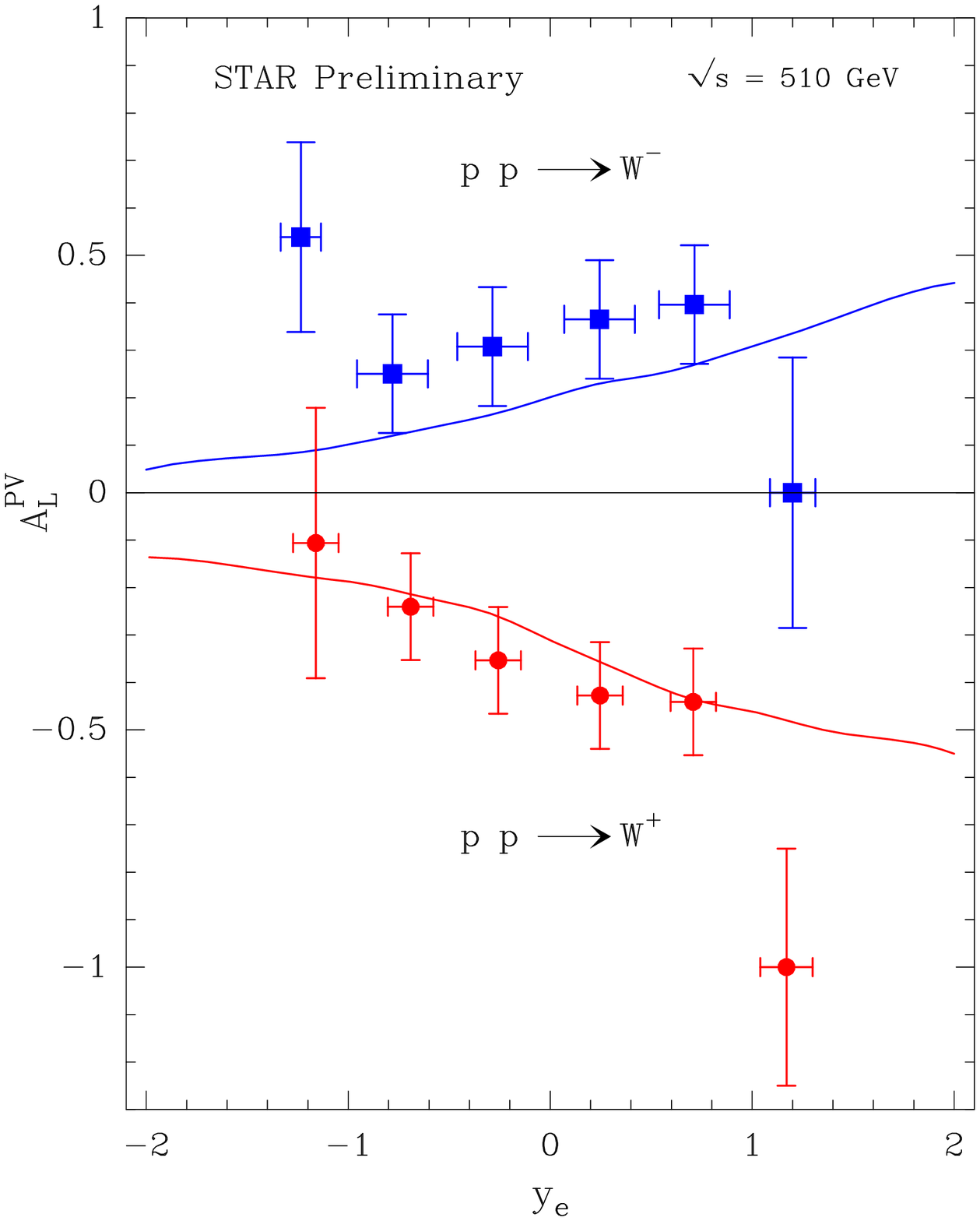}
\caption[*]{\baselineskip 1pt
Our predicted parity-violating helicity asymmetries $A_L^{PV}$ for charged-lepton production at BNL-RHIC, through production and decay of $W^{\pm}$ bosons. $y_e$ is the charged-lepton rapidity and the data points are from Ref. \cite{bs}}
\label{alpv}
\end{center}
\end{figure}

 Let us recall that for the $W^{\pm}$ production in $pp$ collision, the differential cross section $d\sigma_{pp}^{W^{\pm}}/dy$ can be computed directly from the Drell-Yan picture in terms of the {\it dominant} quark-antiquark fusion reactions $u\bar d\to W^+$ and $\bar u d \to W^-$. So for $W^+$ production, we have to lowest-order (LO)
\begin{equation}
d\sigma_{pp}^{W^+}/dy \sim u(x_1, M_{W}^2)\bar {d}(x_2, M_{W}^2) + \bar {d}(x_1, M_{W}^2)u(x_2, M_{W}^2)~,
\label{ppW+}
\end{equation}
where $x_{1,2} = M_W/\sqrt{s} \exp(\pm y)$, $y$ is the rapidity of the W and $\sqrt{s}$ denotes the $c.m.$ energy of the collision. For $W^-$ production, we have a similar expression, after quark flavors interchanged and clearly these $y$ distributions are symmetric under $y \to -y$.\\
In the case of $\bar p p$ collision we have
\begin{equation}
d\sigma_{\bar p p}^{W^+}/dy \sim u(x_1, M_{W}^2) d(x_2, M_{W}^2) + \bar {d}(x_1, M_{W}^2)\bar {u}(x_2, M_{W}^2)~,
\label{barppW+}
\end{equation}
which is no longer symmetric under $y\to -y$, but it simply follows that for $W^-$ production we have $\frac{d\sigma_{\bar p p}^{W^-}}{dy} (y) = \frac{d\sigma_{\bar p p}^{W^+}}{dy} (-y)$.\\
Let us now turn to the charge asymmetry defined as
\begin{equation}
A(y) = \frac{\frac{d\sigma_{\bar p p}^{W^+}}{dy} (y) - \frac{d\sigma_{\bar p p}^{W^-}}{dy} (y)}{\frac{d\sigma_{\bar p p}^{W^+}}{dy} (y) + \frac{d\sigma_{\bar p p}^{W^-}}{dy} (y)}~.
\label{Ay}
\end{equation}
It contains very valuable informations on the light quarks distributions inside the proton and in particular on the ratio of down-to up-quark, as noticed long time ago \cite{bhkw}. Although the cross sections are largely modified by NLO and NNLO QCD corrections, it turns out that these effects do not affect the LO calculation of the charge asymmetry \cite{admp}. A direct measurement of this asymmetry has been achieved by CDF at FNAL-Tevatron \cite{cdf} and the results are shown in Fig. \ref{asym},  together with the prediction of the statistical approach. The agreement is very good and it is remarkable to note that in the high-$y$ region the charge asymmetry tends to
flatten out, following the behavior of our predicted $d(x)/u(x)$ ratio in the high-$x$ region (see Fig. 4 of Ref. \cite{bbs4}).\\
Another interesting case to be studied is the production of $W^{\pm}$ in proton-neutron collision, since this is a realistic future possibility at BNL-RHIC \cite{rikken}. It was first considered theoretically long time ago \cite{bs1} and we recall that for $W^+$ production in $pn$ collision one has 
\begin{equation}
d\sigma_{pn}^{W^+}/dy \sim u(x_1, M_{W}^2)\bar {u}(x_2, M_{W}^2) + \bar {d}(x_1, M_{W}^2)d(x_2, M_{W}^2)~,
\label{pnW+}
\end{equation}
which is not symmetric under $y \to -y$.\\
So here also it is useful to envisage the measurement of the charge asymmetry defined as in Eq.~(\ref{Ay}) and the results of our calculations are shown in Fig. \ref{Rwn}. For both energies there is a rapid rise with the $W$ rapidity and at high-$y$, a similar behavior  to what we have observed for $\bar pp$ collision. Although the charge asymmetry vanishes at $y=0$, there is no reason to expect such a behavior as a function of the lepton rapidity, in the leptonic decay of the $W^{\pm}$ boson. This is an important physical observable that could be measured in the future at RHIC.\\
By combining pp and pn collisions data at BNL-RHIC energies, one can envisage to pin down the flavor asymmetry of the sea from the following charge asymmetry
\begin{equation}
R_{W}(y) = \frac{\frac{d\sigma_{p p}^{W^+}}{dy} (y) - \frac{d\sigma_{p n}^{W^+}}{dy} (y) + \frac{d\sigma_{p p}^{W^-}}{dy} (y) - \frac{d\sigma_{p n}^{W^-}}{dy} (y)}{\frac{d\sigma_{p p}^{W^+}}{dy} (y) + \frac{d\sigma_{p n}^{W^+}}{dy} (y) + \frac{d\sigma_{p p}^{W^-}}{dy} (y) + \frac{d\sigma_{p n}^{W^-}}{dy} (y)}~.
\label{Ry1}
\end{equation}
In terms of parton distributions, it simply reads
\begin{equation}
R_{W}(y) = - \frac{[u(x_1,M_W^2) - d(x_1,M_W^2)][\bar u(x_2,M_W^2) - \bar d(x_2,M_W^2)] + (x_1 \leftrightarrow x_2)}{[u(x_1,M_W^2) + d(x_1,M_W^2)][\bar u(x_2,M_W^2) + \bar d(x_2,M_W^2)] + (x_1 \leftrightarrow x_2)}~.
\label{Ry2}
\end{equation}
Clearly $R_W(y)$ is symmetric under $y \to -y$ and $R_W = 0$ if the sea is flavor symmetric, which is not the case. Since $\bar d(x) >\bar u(x)$, $R_W>0$ and we display in Fig. \ref{RW} our predictions for $R_W(y)$ calculated at two $c.m.$ energies. It is easy to understand the energy behavior of $R_W(y)$, which decreases for increasing energy. For $y=0$ one has $x_1 =x_2$, so when the energy increases this commun value decreases and the difference $\bar u - \bar d$ becomes smaller.\\
Finally let us now turn to spin-dependent observables and more precisely to the parity-violating helicity asymmetries $A_L^{PV}$, which are excellent tools for pinning down the quark helicity distributions, as first noticed in Ref. \cite{bs2}. It allows their flavor separation without the complication of the fragmentation functions one has in semi-inclusive Deep Inelastic Scattering. For a direct determination of $A_L^{PV}(y_W)$ as function of the $W$ rapidity, it was found
that, for $y_W$ large and positive the $W^-$ asymmetry is sensitive to $-\Delta d/d$ and to $\Delta \bar u/\bar u$ for $y_W$ large and negative. Similarly, the $W^+$ asymmetry is sensitive to $\Delta \bar d/\bar d$ for $y_W$ large and positive and to $-\Delta u/u$ for $y_W$ large and negative. However unlike for the charge asymmetry considered above from the CDF data, it is not possible to get a direct reconstruction of the $W$ helicity asymmetry at BNL-RHIC and therefore we must cope with the helicity asymmetry at the lepton level.\\
In this case we consider the processes $\overrightarrow p p\to W^{\pm} + X \to e^{\pm} + X$, where the arrow denotes a longitudinally polarized proton and the outgoing $e^{\pm}$ have been produced by the leptonic decay of the $W^{\pm}$ boson. The helicity asymmetry is defined as
\begin{equation}
A_L^{PV} = \frac{d\sigma_+ - d\sigma_-}{d\sigma_+  + d\sigma_-}~.
\label{AL}
\end{equation}
Here $\sigma_h$ denotes the cross section where the initial proton has helicity $h$.\\ For $W^-$ production, the numerator of the asymmetry is found to be proportional to
\begin{equation}
\Delta \bar u(x_1,M_W^2)d(x_2,M_W^2)(1 - \mbox{cos} \theta)^2 - \Delta d(x_1,M_W^2)\bar u(x_2,M_W^2)(1 + \mbox{cos} \theta)^2~,
\label{AL1}
\end{equation}
where $\theta$ is the polar angle of the electron in the $c.m.s.$, with $\theta =0$ in the forward direction of the polarized parton. The denominator of the asymmetry
has a similar form, where the quark helicity distributions are replaced by unpolarized ones and with a plus sign between the two terms of the above expression. For $W^+$ production, the asymmetry is obtained by interchanging the quark flavors ($ u \leftrightarrow d$).\\ We first show in Fig. \ref{alpv} the results of the calculations \footnote{We are grateful to Prof. W. Vogelsang for performing this numerical work. A similar calculation using the program RHICBOS \cite{ny} was presented earlier \cite{js}.} of the helicity asymmetries, versus the charged-lepton rapidity $y_e$, but for a clear interpretation some explanations are required \cite{fv}. At high negative $y_e$, one has $x_2 >>x_1$ and $\theta >>\pi/2$, so the first term in Eq. (\ref{AL1}) dominates and the asymmetry generated by the $W^-$ production is driven by $\Delta \bar u(x_1)/\bar u(x_1)$, for medium values of $x_1$. Similarly for high positive $y_e$, the second term
in Eq. (\ref{AL1}) dominates and now the asymmetry is driven by $-\Delta d(x_1)/d(x_1)$, for large values of $x_1$. So we have a clear separation between these two contributions, like in the hypothetic case of the reconstructed $W^-$ asymmetry. We note in Fig. \ref{alpv} that the $W^-$ asymmetry remains positive at high positive $y_e$, which reflects the fact that $\Delta d(x)/d(x)$ remains negative at high-$x$, as predicted by the statistical approach (see Fig. 5 of Ref. \cite{bbs4}). For the helicity asymmetry generated by the $W^+$ production, by examining the expression corresponding to Eq. (\ref{AL1}), one finds a different situation because both terms compete and it does not allow a clean kinematical separation, between $\Delta \bar d/\bar d$ and $-\Delta u/u$ \footnote{  This separation can be obtained by measuring the $W^-$ asymmetry in $p\overrightarrow n$ collision, with polarized He-3, as noticed in Ref. \cite{rikken}.}. However one observes that the $W^+$ asymmetry is negative, which is most probably due to the fact that $\Delta u$ is positive and large. Finally, our predictions in Fig. \ref{alpv} seem to be in fair agreement with some preliminary data from STAR at BNL-RHIC, presented recently \cite{bs}.\\

Therefore the continuous agreement of our predictions with existing data leads us to conclude that our physical approach is based on very solid grounds.

\clearpage
\newpage

\clearpage
\newpage

\clearpage
\newpage

\clearpage
\newpage


\begin{thebibliography}{99}


\bibitem{bbs1} C. Bourrely, F. Buccella and J. Soffer, Eur. Phys. J. C 23 (2002) 487.

\bibitem{bbs2} C. Bourrely, F. Buccella and J. Soffer, Phys. Lett. B 648 (2007) 39.

\bibitem{bbs3}  C. Bourrely, F. Buccella and J. Soffer, Mod. Phys. Lett. A 18 (2003) 771.

\bibitem{bbs4}  C. Bourrely, F. Buccella and J. Soffer, Eur. Phys. J. C 41 (2005) 327.

\bibitem{bbs5} C. Bourrely, F. Buccella and J. Soffer, Phys. Rev. D 83 (2011) 074008.

\bibitem{bbs6} C. Bourrely, F. Buccella and J. Soffer, Int. J. of Mod. Phys. A 28 (2013) 1350026.

\bibitem{bhkw} E.L. Berger, F. Halzen, C.S. Kim and S. Willenbrock, Phys. Rev. D 40 (1989) 83; Erratum, ibid 3789.

\bibitem{admp} Ch. Anastasiou, L. Dixon, K. Melnikov and F. Petriello, Phys. Rev. D 69 (2004) 094008.

\bibitem{cdf} T. Aaltonen {\it et al.} (CDF Collaboration), Phys. Rev. Lett. 102 (2009) 181801.

\bibitem{rikken} Workshop on Opportunities for Polarized He-3 in RHIC and EIC, BNL Upton, NY, USA, September 28-30, 2011.

\bibitem{bs1} C. Bourrely and J. Soffer, Nucl. Phys. B 423 (1994) 329.

\bibitem{bs2} C. Bourrely and J. Soffer, Phys. Lett. B 314 (1993) 132.

\bibitem{ny} P. Nadolsky and C.P. Yuan,  Nucl. Phys. B 666 (2003) 31 (see http://hep.pa.msu.edu/resum/).

\bibitem{js} J. Soffer, Invited talk at "`Parity-Violating Spin Asymmetries at RHIC-BNL"', April 26-27, 2007, Proceedings BNL-79146,  Editors, M. Grosse Perdekamp, B. Surrow and W. Vogelsang.

\bibitem{fv} D. De Florian and W. Vogelsang, Phys. Rev. D 81 (2010) 094020.


\bibitem{bs} B. Surrow, Contribution at the XXI Int. Workshop on DIS and Related Subjects-DIS13, Marseille, France, April 22-26, 2013.

\end{thebibliography}
\end{document}